\documentclass[doublecol]{epl2} 
\usepackage{ulem}
\usepackage{verbatim} 

\usepackage[utf8]{inputenc}
\usepackage{amsmath}
\usepackage{booktabs}
\usepackage{amssymb}

\begin{document}
\title{A minimal model for the Weyl nodes and Fermi arcs of PtBi$_2$}

\author{Tobias Cristófoli\inst{1}\and Manuel Alonso Lemos\inst{1} \and Jorge I. Facio\inst{1,2,3} \and Pablo S. Cornaglia\inst{1,2,3}}
\shortauthor{T. Cristófoli \etal}

\institute{
\inst{1}Centro Atómico Bariloche and Instituto Balseiro, CNEA, 8400 Bariloche, Argentina\\
\inst{2}Consejo Nacional de Investigaciones Científicas y Técnicas (CONICET), Argentina\\
\inst{3}Instituto de Nanociencia y Nanotecnología CNEA-CONICET, Argentina}


\abstract{
Weyl semimetals host topologically protected Fermi arcs on their surfaces, originating from the Chern number of the bulk Weyl nodes. 
In trigonal PtBi$_2$, superconducting signatures have been associated with the Fermi arcs.
Theoretical descriptions of this surface superconductivity have so far relied on effective models that are not directly tied to the microscopic electronic structure of the material. In this work we develop a minimal description of the low-energy bands guided by density functional theory calculations and fully constrained by the relevant crystalline and time-reversal symmetries. The model captures the evolution of the Weyl nodes with spin--orbit coupling, including node annihilations, and is able to describe the spin-momentum locking of the surface Fermi arcs. It reproduces the orbital content and the band topology of PtBi$_2$ and provides a starting point for further studies.}

\maketitle

Trigonal PtBi$_2$ (t-PtBi$_2$) has attracted considerable attention as a material in which lattice, orbital, and spin degrees of freedom are all simultaneously active: a low-symmetry crystal structure, several Bi $6p$ and Pt $5d$ orbitals close to the Fermi level, and the strong spin-orbit coupling (SOC) characteristic of heavy elements~\cite{10.1063/5.0272618,feng_rashba-like_2019}. One of its interesting aspects is that the system realizes a Weyl semimetallic phase~\cite{Veyrat2023}, the associated topological Fermi arcs have been measured by different techniques~\cite{Kuibarov2023,hoffmann2024fermi,n5pz-j2sl,changdar2025topological,mathisen2026fermiologyspinpolarizationtopological} and, remarkably, it has been indicated that the Fermi arcs host superconductivity~\cite{Kuibarov2023,changdar2025topological,kuibarov_measuring_2025}.

These findings are naturally fueling further experimental and theoretical studies of both the normal~\cite{PhysRevB.110.054504,PhysRevB.110.125148,PhysRevResearch.7.013025,palumbo2025gapless,palumbo2025interplay,f66s-m6jy} and superconducting phases~\cite{schimmel2024surface,Zabala_2024,bdtb-mb8c,kkqg-ntcz,47vs-qgzk,besproswanny2025,maeland2025mechanism,buccheri2026,vocaturo2026}. At the same time, the same richness of lattice, orbital, and spin degrees of freedom that makes the material attractive also works against a simple understanding of its Weyl phase: realistic first-principles descriptions involve many bands, and it is not obvious which physical degrees of freedom are essential to the low-energy Weyl physics near the Fermi level and which can be discarded~\cite{PhysRevB.110.054504}.

In this work, we develop a minimal symmetry-based model for trigonal PtBi$_2$ that bridges the gap between microscopic calculations and phenomenological descriptions. Guided by \textit{ab initio} results, we identify the essential orbital degrees of freedom underlying the Weyl nodes and construct a continuum Hamiltonian constrained by the crystal symmetries and time-reversal invariance. The model reproduces the topology of the Weyl phase, the SOC-driven evolution of the Weyl nodes, and the spin texture of the surface Fermi arcs. It thus provides a transparent framework for investigating the role of Fermi arcs in the anomalous surface superconductivity reported in PtBi$_2$.

The crystal structure of trigonal PtBi$_2$ is composed of stacked layers of Pt and Bi, each a distorted triangular lattice with three atoms per unit cell~\cite{ptbi2:kaiser14,Shipunov2020}. Our starting point is set by previous density functional theory (DFT) calculations, which reveal a set of 12 Weyl nodes close to the Fermi level~\cite{Veyrat2023}. A priori, the Bloch states forming these nodes are distributed over several degrees of freedom---the electron spin, the Bi $6p$ and Pt $5d$ orbitals, and the six Bi and three Pt sites of the unit cell. Building a tractable low-energy model therefore requires reducing this large basis to the few degrees of freedom that actually carry the topology of the nodes.

\begin{figure}
    \centering
    \includegraphics[width=\linewidth]{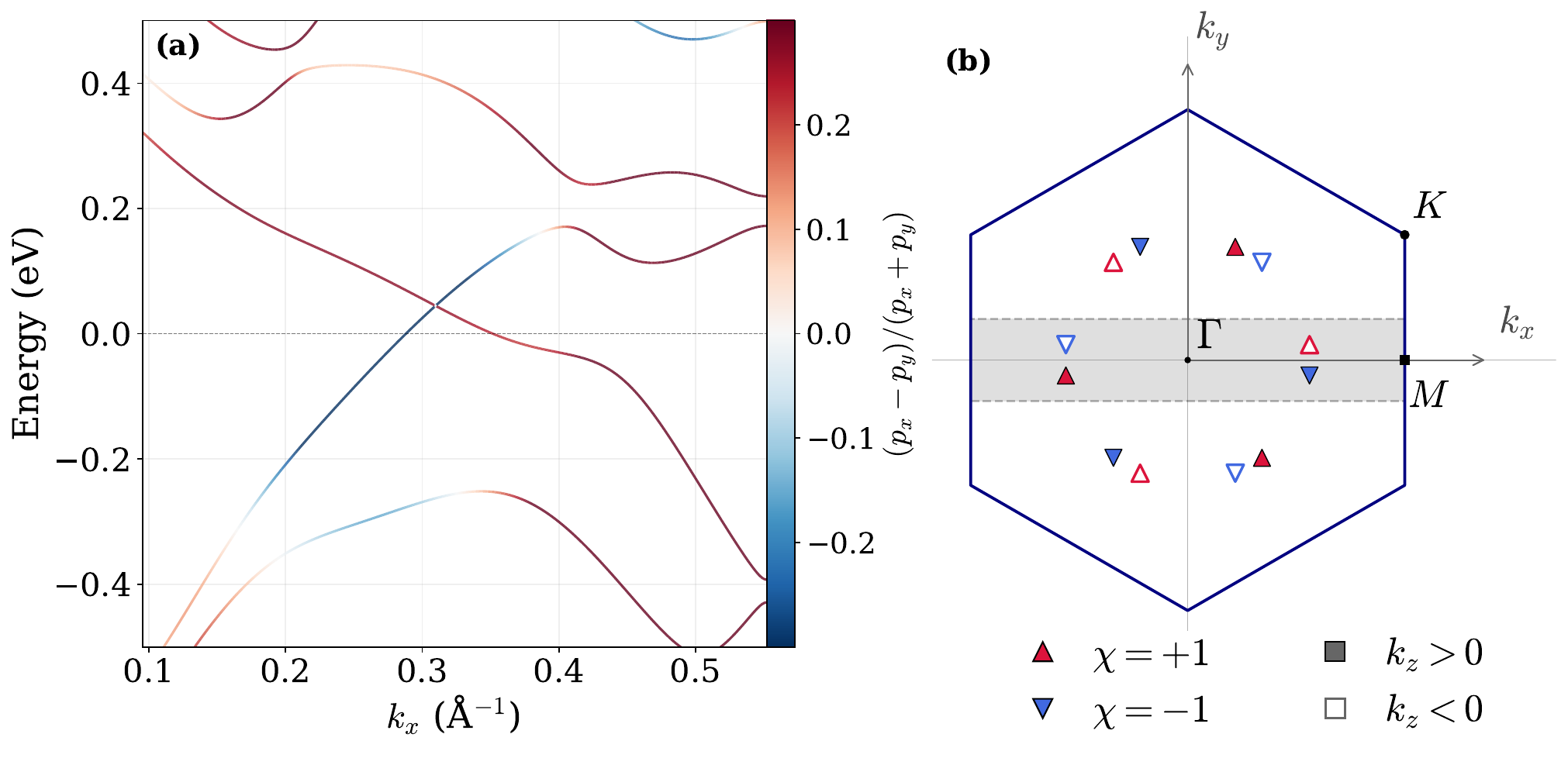}
\caption{
(a) Calculated electronic band structure in the vicinity of a Weyl node. The color scale indicates the weight of the Bi $6p_x$ and $6p_y$ orbitals, showing that the bands forming the Weyl node are predominantly of in-plane Bi $6p$ character. 
(b) Weyl points (WPs) in the Brillouin zone (top view).
The twelve nodes form a single orbit of
$\mathbf{w}=(0.324,\,0.041,\,-0.152)\,2\pi/a$ under the
$C_{3v}$ point group and time reversal $\mathcal{T}$.
The threefold rotation $C_3$ and $\mathcal{T}$ ($\mathbf{k}\!\to\!-\mathbf{k}$)
connect WPs of equal chirality
(red $\blacktriangle$, $\chi=+1$; blue $\blacktriangledown$, $\chi=-1$),
whereas the mirror planes $\sigma_v$ (one acting as $k_y\!\to\!-k_y$)
connect WPs of opposite chirality.
Filled/open markers denote $k_z\gtrless 0$.
The shaded strip spans the zone along $k_x$ and encloses the two WP pairs nearest to the $k_x$ axis.}
\label{fig:DFTpxpy}
\end{figure}

A non-vanishing Chern number means that the cell-periodic part $u_{n\mathbf{k}}$ of the Bloch state cannot be chosen smoothly on a surface enclosing the node, so that some combination of the microscopic degrees of freedom encoded in $u_{n\mathbf{k}}$ must change rapidly along any path around
 it. While the complex crystal structure of trigonal PtBi$_2$ makes the identification of the corresponding low-energy pseudospin challenging, the DFT calculations reveal that the Bi $\{6p_x,6p_y\}$ orbitals provide the leading contribution to the variation of the Bloch states across the nodes. This is illustrated in Fig.~\ref{fig:DFTpxpy} for one of the nodes near the $\Gamma$--$M$ line as a function of $k_x$. While other orbitals, such as Bi $6p_z$ or Pt $5d$, carry a non-negligible weight on these states, they remain largely spectators of the Weyl physics, their contribution varying little across the node (see Supplementary Material, SM). In light of this, we adopt a minimalist approach by retaining only two effective orbital levels representing the $p_x$ and $p_y$ states.

A second difficulty  stems from the multiplicity of nodes imposed by the symmetries. One possibility is to enforce these symmetries within a tight-binding model. However, even after restricting to the $\{p_x,p_y\}$ orbitals, a description retaining all six Bi sites would still lead to an unwieldy model. Moreover, the low-energy physics is intimately tied to the structural distortion that lowers the translational symmetry relative to the parent triangular lattice~\cite{palumbo2025gapless,palumbo2025interplay}; we therefore choose not to work with an effective triangular-lattice model, but rather in the continuum approximation. 

In this approach, much as the two valleys of graphene are treated as independent low-energy sectors, we group the nodes into \textit{sectors}: each sector defines an independent block of the Hamiltonian, and the full model is  assembled by combining the blocks generated by the remaining symmetries. The grouping is not unique, but it is constrained by a physical requirement: since we  also aim to describe the experimentally observed Fermi arcs, we demand that nodes linked by a Fermi arc belong to the same sector. The observed arcs connect nodes whose projections onto the $(001)$ surface lie close together along $\Gamma$--$M$; these are pairs of nodes related by a  mirror times time reversal, such as $M_x\mathcal{T}$. In the main text, we take a sector, which we call sector~0, comprising the four such nodes nearest the $\Gamma$--$M$ line, connected by the mirror $M_x$ and time reversal $\mathcal{T}$ and enclosed by the shaded rectangle in Fig.~\ref{fig:DFTpxpy}; the remaining $C_3$-related nodes are then introduced as additional blocks within a valley approach. In the SM, we adopt a finer partition, based on an elementary valley containing a single arc-connected pair of nodes.

\section{The model} 
We start without SOC, in which case DFT calculations in PtBi$_2$ indicate the existence of Weyl nodes in each spin projection~\cite{palumbo2025gapless,palumbo2025interplay}.
In this limit, the effective Hamiltonian in sector 0 is written in the basis of Pauli matrices $\tau_i$ acting on the orbital space:
\begin{equation}
H_0(\mathbf{k}) = \mathbf{f(k)}\cdot \boldsymbol{\tau}. 
\end{equation} 
Here $\boldsymbol{\tau}=(I,\tau_x,\tau_y,\tau_z)$, where $I$ is the $2\times 2$ identity matrix and $\tau_{i}$ the Pauli matrices acting on the orbital basis $\{p_x, p_y\}$.

The form of each $f_i(\mathbf{k})$ is dictated by the constraints of mirror symmetry $M_x$ and time reversal $\mathcal{T}$. The mirror acts as $U_{M_x} = -\tau_z$, while time reversal for real orbitals imposes $H_0(\mathbf{k})^* = H_0(-\mathbf{k})$. These symmetries require that $f_3$ be even in $k_x$ and even under $(k_y, k_z) \to (-k_y, -k_z)$; that $f_1$ be odd in $k_x$ and odd under $(k_y, k_z) \to (-k_y, -k_z)$; and that $f_2$ be odd in $k_x$ but even under $(k_y, k_z) \to (-k_y, -k_z)$. The expressions in Eqs.~(2)--(4) represent the lowest-order symmetry-allowed polynomials\footnote{The symmetry allowed terms were obtained using \texttt{qsymm} \cite{varjas2018qsymm} and checked numerically.} satisfying these parity rules that simultaneously vanish at a common point $\mathbf{w} = (w_x, w_y, w_z)$, thereby guaranteeing Weyl nodes at this location and at its symmetry-related images.

\begin{align}
    f_1(\tilde{\mathbf{k}}) &= a_1\tilde{k}_x(\tilde{k}_z - \tilde{k}_y),\\
    f_2(\tilde{\mathbf{k}}) &= a_2\tilde{k}_x \left(1-\tilde{k}_z^2\right),\\
    f_3(\tilde{\mathbf{k}}) &= a_3\left(1-\tilde{k}_x^2\right),
\end{align}
where, $\tilde{k}_\alpha=\frac{k_\alpha}{w_\alpha}$ is the dimensionless momentum component. 
This model yields 4 isolated Weyl nodes at $\mathbf{w}=\{w_x, w_y, w_z\}$ and their symmetry-related positions.
The corresponding velocity matrix is given by $v_{ij}=\partial f_i/\partial k_j\big|_{\mathbf{k}=\mathbf{w}}$,
and the chirality $\chi=\text{sign}(\det v)=-\text{sign}(a_1 a_2 a_3/w_x w_y w_z)$. 
The $a_i$ parameters control the nonzero components of the velocity at the node, while the $w_i$ set the node positions directly. 
Reproducing the full velocity matrix from the DFT band structure, including the tilt of the Weyl cones, which we set to zero through $f_0(\mathbf{k})=0$, would require additional symmetry-allowed terms that we omit for simplicity. 
A finite $f_0$ can be introduced if needed, as it affects neither the position nor the chirality of the nodes.

For a slab spanning the $x$--$y$ plane, this model leads to Fermi arcs on the surfaces of the slab. The arcs join pairs of Weyl node projections on the $k_x$--$k_y$ plane  $(w_x,\pm w_y)$, and $(-w_x,\pm w_y)$ with a straight line.

In order to introduce a dispersion on the Fermi arc we included a symmetry allowed term that does not change the chirality or position of the nodes $\mu\left(1-\tilde{k}_y^2\right)\tau_z$. When $1 + a_3/\mu > 0$, however, this generates an accidental nodal line at $k_x = 0$ that can be gapped adding a term $\beta(1-\tilde{k}_y^4)\tau_z$, which results in:\footnote{See SM for the allowed values of $\beta$.}
\begin{equation}
       f_3(\tilde{\mathbf{k}}) = a_3\left(1-\tilde{k}_x^2\right)+ \mu\left(1-\tilde{k}_y^2\right)+ \beta\left(1-\tilde{k}_y^4\right).
\end{equation}

\textit{Spin-orbit coupling.} 
When spin is included, the basis becomes $(\uparrow,\downarrow)\otimes(p_x,p_y)$ and the Hamiltonian is a $4\times 4$ matrix. In the absence of SOC, the Weyl nodes are doubly degenerate and each one has an associated topological charge of $\pm 2$.
DFT results show that gradually turning on the SOC coupling leads to a splitting of each of the Weyl nodes into two nodes with the same chirality. The resulting nodes move through the Brillouin zone and at the physically relevant SOC coupling intensity, Weyl nodes having opposite chirality annihilate in pairs on the $k_z=k_y=0$ axis, and the symmetry related planes~\cite{palumbo2025interplay}. 

We now include a SOC term in our minimal model in order to describe this behavior. Time reversal acts as $\mathcal{T}=i\sigma_y K\otimes I_\mathrm{orb}$ and the mirror as $M_x = -i\sigma_x\otimes\tau_z$, both satisfying $\mathcal{T}^2=M_x^2=-1$ as required for spin-$\tfrac{1}{2}$. The symmetry-allowed SOC terms follow from imposing both constraints simultaneously, using the action of $\mathcal{T}$ and $M_x$ on the spin and orbital matrices detailed in the SM.
 
A general expansion up to linear order in $\mathbf{k}$ yields fifteen symmetry-allowed SOC terms.
Among these, we first consider the two that are momentum-independent:
\begin{equation}
    H_\mathrm{SOC} = g_1\,\sigma_z\otimes\tau_y + g_2\,\sigma_y\otimes\tau_y.
\end{equation}
Being momentum-independent, these terms correspond to 
purely on-site interactions in real space, and can be identified as components 
of the atomic SOC
projected onto the $\{p_x, p_y\}$ subspace, where the orbital matrix $\tau_y$ acts as 
$L_z$. Since the dominant SOC in heavy elements such as Bi is 
the intra-atomic term, we focus on this contribution as the primary driver of 
the Weyl node annihilation, and treat the remaining momentum-dependent terms 
as perturbative corrections that govern the spin texture of the surface states 
without altering the node count at leading order.

Both momentum-independent SOC terms share the same orbital matrix $\tau_y$ and can be written compactly as
\begin{equation}
    H_\mathrm{SOC} = g\,(\hat{n}\cdot\vec{\sigma})\otimes\tau_y,
\end{equation}
where $g = \sqrt{g_1^2+g_2^2}$ is the effective SOC strength and $\hat{n}=(0,g_2,g_1)/g$ defines the spin quantization axis, tilted in the $yz$-plane. Because $\hat{n}\cdot\vec{\sigma}\otimes I$ commutes with the full Hamiltonian $H = \sigma_0\otimes\mathbf{f}(\mathbf{k})\cdot\boldsymbol{\tau} + H_\mathrm{SOC}$, the latter splits exactly into two $2\times 2$ blocks labeled by the eigenvalue $s=\pm 1$ of $\hat{n}\cdot\vec{\sigma}$:
\begin{equation}
    H_s(\mathbf{k}) = f_1\,\tau_x + (f_2 + s g)\,\tau_y + f_3\,\tau_z.
\end{equation}
The exact eigenvalues of the full $4\times 4$ Hamiltonian are
\begin{equation}
\varepsilon_s = \pm\sqrt{f_1^2 + (f_2+s g)^2 + f_3^2}.
\end{equation}
A Weyl node occurs when $\varepsilon_s=0$ in one of the two sectors, which requires
\begin{equation}
    f_1 = 0,\qquad f_2 = -s\, g,\qquad f_3 = 0.
\end{equation}
From $f_1=0$ with $k_x\neq 0$ one immediately obtains $k_y/w_y = k_z/w_z$, so all nodes are constrained to a plane in momentum space.

At $g=0$, each spinless node is doubly degenerate due to the trivial spin degeneracy. A finite $g$ lifts this degeneracy: each doubly-degenerate point splits into a pair of Weyl nodes belonging to opposite sectors of $\hat{n}\cdot\vec{\sigma}$. As $g$ increases, two nodes from different original pairs migrate toward each other. At a critical value $g^c$, two pairs with opposite chirality meet and annihilate at $k_y = k_z = 0$. Because $f_2 = a_2(k_x/w_x)$ at this axis, the collision occurs at a finite momentum $k_x = -s g^c w_x / a_2$. 
Of the two pairs in each spin sector, one has transverse coordinate
$\tilde{k}_y=\tilde{k}_z=\pm\sqrt{1+g/|a_2|}$ that grows with $g$ and
survives, while the partner pair, with $\tilde{k}_y^2 = 1-g/|a_2|$,
shrinks and annihilates when $\tilde{k}_y\to0$. Setting
$\tilde{k}_y=\tilde{k}_z=0$, $f_2=-sg$ gives $\tilde{k}_x=-sg/a_2$ and
$f_3=0$ gives $\tilde{k}_x^2=1+(\mu+\beta)/a_3$; squaring the former and
eliminating $\tilde{k}_x$,
\begin{equation}
    g^c = |a_2|\sqrt{1+\frac{\mu+\beta}{a_3}},
    \label{eq:gc}
\end{equation}
which is independent of $s$, so both spin sectors annihilate at the same
threshold. For ($|\mu|,|\beta| \ll |a_3|$), we have $g^c\simeq|a_2|$.

Beyond $g^c$, only two pairs of Weyl nodes survive. The conditions $f_1=0$, $f_2=-sg$, $f_3=0$ admit a closed-form solution. The constraint $f_1=0$ (with $\tilde{k}_x\neq 0$) imposes $\tilde{k}_y = \tilde{k}_z$, and substitution into $f_3=0$ reduces the system to a single quartic equation for $\tilde{k}_x$.

To first order in $\mu/a_3$, the surviving pair is located at
\begin{align}\label{eq:traj}
    \tilde{k}_x &\approx s\,\mathrm{sgn}(a_2) \left(1 - \frac{(\mu+2\beta)\,g}{2\,|a_2|\, a_3}
    - \frac{\beta g^2}{2\,a_2^2 a_3}\right),\\
    \tilde{k}_y &= \tilde{k}_z = \pm\sqrt{1 + g/|a_2|},\nonumber
\end{align}
with $s=\pm 1$ the spin sector; for the parameters used here ($a_2<0$) the sector-$s$ pair thus sits at $\tilde{k}_x\simeq -s$.

Importantly, while the topology (number and position of the Weyl nodes) is controlled entirely by the magnitude $g$, the individual values of $g_1$ and $g_2$ determine the spin-locking axis $\hat{n}$ of the Fermi arcs, rotating it away from $\hat{z}$ toward $\hat{y}$ by an angle $\theta = \arctan(g_2/g_1)$.
The momentum-dependent terms do not change the topology at leading order, but they control the tilt of the spin orientation away from $\hat{n}$ and the spin-momentum texture on the Fermi arcs.

The linear-in-k symmetry-allowed SOC couplings generally mix the two $\hat{n}\cdot\vec{\sigma}$ sectors and break the exact block-diagonal structure of $H$.\footnote{See SM for the full list of symmetry allowed terms.} We focus here on two representative couplings.
First, we consider
\begin{equation}
    V_3 = g_3\,\tilde{k}_x\,\sigma_x\otimes\tau_x,
\end{equation}
and treat its effect on the node positions perturbatively (see SM). This term does not change the geometric constraints $f_1=f_3=0$ but acts purely along the $\tau_y$ axis at second order, renormalising the on-site SOC to an effective magnitude
\begin{equation}
    g \;\longrightarrow\; g_\mathrm{eff} \;=\; g - \frac{g_3^2}{2g} \;<\; g,
    \label{eq:g_eff}
\end{equation}
evaluated at the unperturbed nodes ($\tilde{k}_x^2\approx 1$). The surviving Weyl pairs therefore retreat backward along the trajectory of Eq.~(\ref{eq:traj}) under the substitution $g\to g_\mathrm{eff}$, and the annihilation threshold is pushed to the renormalised value $g^c \approx |a_2| + g_3^2/(2 |a_2|)$. The average spin acquires a $k_x$ dependence which is also present on the Fermi-arc states.

A second symmetry-allowed coupling provides a complementary route to control the spin texture of the surface states,
\begin{equation}
    V_4 = g_4\,\tilde{k}_y\,\sigma_y\otimes\tau_x.
    \label{eq:V4}
\end{equation}
Unlike $V_3$, which is purely off-diagonal in the $\hat{n}\cdot\vec{\sigma}$
basis, $\sigma_y$ has a finite projection on
$\hat{n}=(0,g_2,g_1)/g$. Decomposing
$\sigma_y = (g_2/g)\,\hat{n}\cdot\vec{\sigma} + \sigma_y^{\perp}$, the parallel
piece acts diagonally in each spin sector and renormalises the $\tau_x$
coefficient of the block Hamiltonian at first order,
\begin{equation}
    f_1(\mathbf{k}) \;\longrightarrow\; f_1(\mathbf{k}) + s\,g_4\,\tilde{k}_y\,\frac{g_2}{g},
    \label{eq:f1_shift}
\end{equation}
lifting the constraint $\tilde{k}_y=\tilde{k}_z$ that pinned the surviving Weyl
pairs to a single plane in the $V_3$ analysis. The transverse component
$\sigma_y^{\perp}$ leads to a further renormalisation of the effective $g$ parameter.

The two contributions of
$V_4$ are weighted by complementary projections of the on-site SOC axis:
$g_2/g$ sets the first-order shift of $f_1$, while
$g_1/g$ sets the second-order renormalisation of $g$. 
The full momentum-dependent SOC,
$V_3+V_4 = (g_3\tilde{k}_x\sigma_x + g_4\tilde{k}_y\sigma_y)\otimes\tau_x$,
realises a generic in-plane spin-momentum locking in the orbital-flip
channel.  While the bulk node topology remains controlled by $g_\mathrm{eff}$, the average direction of the
spins can be modified through the tilt of $\hat{n}$,  $g_3$, and $g_4$.

The effect of any linear-in-$k$ SOC term $\sigma_i\otimes\tau_j$ on both the node positions and the bulk spin texture  can be understood by decomposing its spin operator relative to the on-site axis $\hat{\mathbf n}=(0,g_2,g_1)/g$. 
The longitudinal part of $\sigma_i$ ($\parallel\hat{\mathbf n}$) commutes with $\hat{\mathbf n}\cdot\vec{\sigma}$ and acts within each spin block at first order, shifting a block field, the orbital factor selecting $f_1$ for
$\tau_x$, $f_3$ for $\tau_z$, and a cone tilt for $\tau_0$, and so displaces the nodes while leaving the spin along $\pm\hat{\mathbf n}$. 
The transverse part ($\perp\hat{\mathbf n}$) instead couples the two blocks, with two different
effects: at first order it cants the spin away from $\hat{\mathbf n}$ (the spin texture, analyzed below), and at second order it renormalises the on-site SOC $g\to g_\mathrm{eff}$, displacing the nodes. Because
$\hat{\mathbf n}$ lies in the $yz$ plane, $\sigma_x$ is purely transverse,
whereas $\sigma_y,\sigma_z$ enter the first-order (node) shift weighted by
$g_2/g$ and $g_1/g$. 
These effects are summarized in Table~\ref{tab:soc_effects}, with the explicit $g_\mathrm{eff}$ given in the
SM; $V_3$ and $V_4$ are the two representative cases, $g_3$ only renormalising $g$ while $g_4$ additionally shifts $f_1$.

\begin{figure}[t]
  \centering
  \includegraphics[width=\columnwidth]{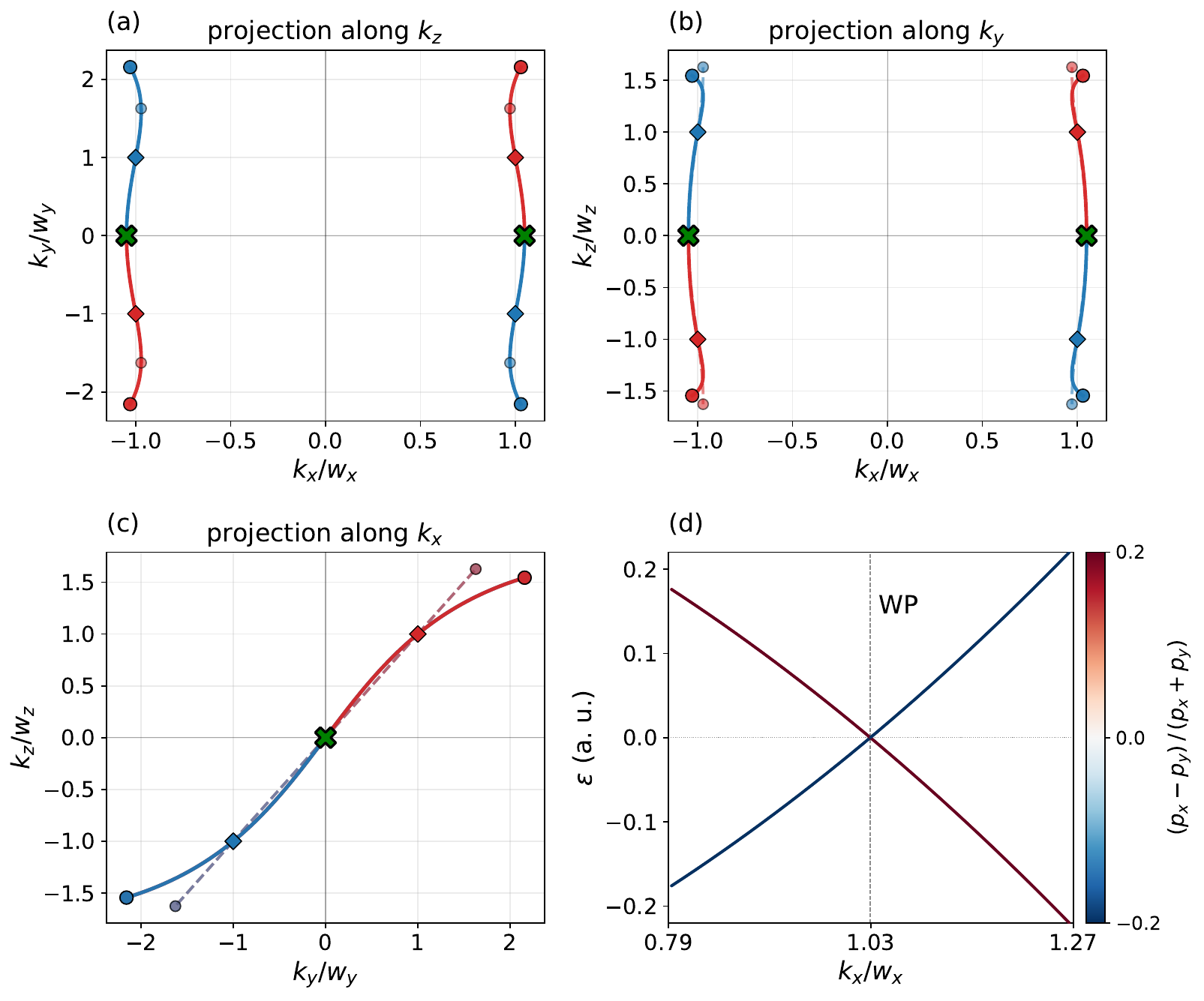}
  \caption{ 
    (a)--(c) Weyl-node trajectories in momentum space as the SOC coupling intensity $\lambda$ is increased:
    (a) projection along $k_z$.
    (b) projection along $k_y$, and
    (c) projection along $k_x$.
    Solid lines correspond to the parameters in Table \ref{tab:params} and dashed lines to $g_3=g_4=0$; red and blue indicate the chirality $\chi=+1$ and
    $\chi=-1$, respectively. Rhombuses mark the spinless starting points
    ($\lambda=0$), filled circles correspond to nodes surviving at $\lambda=1$, and
    green crosses the pairwise annihilation events.
    (d) Electronic band structure along $k_x/w_x$ through a Weyl
    node at full SOC. The color scale indicates the weight of the $p_x$ and $p_y$ orbitals.}
  \label{fig:weyl_annihilation}
\end{figure}

To illustrate the node annihilation as the SOC coupling is turned on, we considered the set of model parameters listed in Table \ref{tab:params}. The spinless node positions and the $a_i$ parameters were obtained by performing a least-squares-fit of the DFT bands near one of the Weyl nodes for zero SOC (see Ref. \cite{palumbo2025interplay} for the DFT calculation details). In order to drive the node annihilation, we first considered the case with only the two coupling terms $g_1$ and $g_2$, and used a dimensionless parameter $\lambda\in[0,1]$ to interpolate linearly between the spinless model ($\lambda=0$) and the full SOC ($\lambda=1$) case. The value of $g=\sqrt{g_1^2+g_2^2}=0.16\,eV$ exceeds the annihilation threshold value $\sim |a_2|=0.1 eV$. In normalized coordinates ($k_\alpha/w_\alpha$) the nodes start for $\lambda=0$ at $\tilde{k}=(1,1, 1)$ and at $\mathcal {T}$, $M_x$ related positions. These starting positions are indicated with rhombuses in Fig. \ref{fig:weyl_annihilation}. As the SOC is turned on, each of these nodes split into two nodes having the same chirality that move in different $k$ directions. Two pairs of nodes having opposite chirality annihilate at $k_y=k_z=0$ and $\tilde{k}_x\sim \pm 1$. The nodes are constrained to the $k_y=k_z$ plane as expected. Introducing $g_3$ and $g_4$ does not change the annihilation mechanism as the four SOC couplings $(g_1,g_2,g_3,g_4)$ are increased linearly. The $k_y=k_z$ constraint is however removed by the $g_4$ coupling as discussed above. 

Figure \ref{fig:weyl_annihilation}(d) also shows the bands as a function of $k_x$ colored by the orbital character with all SOC of Table \ref{tab:params} parameters turned on. This behavior mimics the one observed in the DFT results (see Fig \ref{fig:DFTpxpy}).

\begin{table}[htpb]
\centering
\caption{Model parameters.}
\label{tab:params}
\renewcommand{\arraystretch}{1.3}
\begin{tabular}{@{} l l l l @{}}
\toprule
\multicolumn{4}{@{}l}{\textit{Weyl-node position for the spinless model} (1/\AA)} \\[2pt]
$w_x{:}\ 0.4242$ & $w_y{:}\ {-0.041}$ & $w_z{:}\ 0.123$ \\
\midrule
\multicolumn{4}{@{}l}{\textit{Spinless model} ($eV$)} \\[2pt]
$a_1{:}\ -0.15$ & $a_2{:}\ -0.1$ & $a_3{:}\ {0.4}$ \\
\midrule
\multicolumn{4}{@{}l}{\textit{Arc shape} ($eV$)}\\[2pt]
$\mu{:}\ {0.05}$ & $\beta{:}\ -0.01$ & \\
\midrule
\multicolumn{4}{@{}l}{\textit{Spin--orbit coupling} ($eV$)} \\[2pt]
$g_1{:}\ 0.1$ & $g_2{:}\ -0.125$ & $g_3{:}\ 0.05$ & $g_4{:}\ -0.05$ \\
\bottomrule
\end{tabular}
\end{table}

\begin{figure*}[t]
  \centering
  \includegraphics[width=\textwidth]{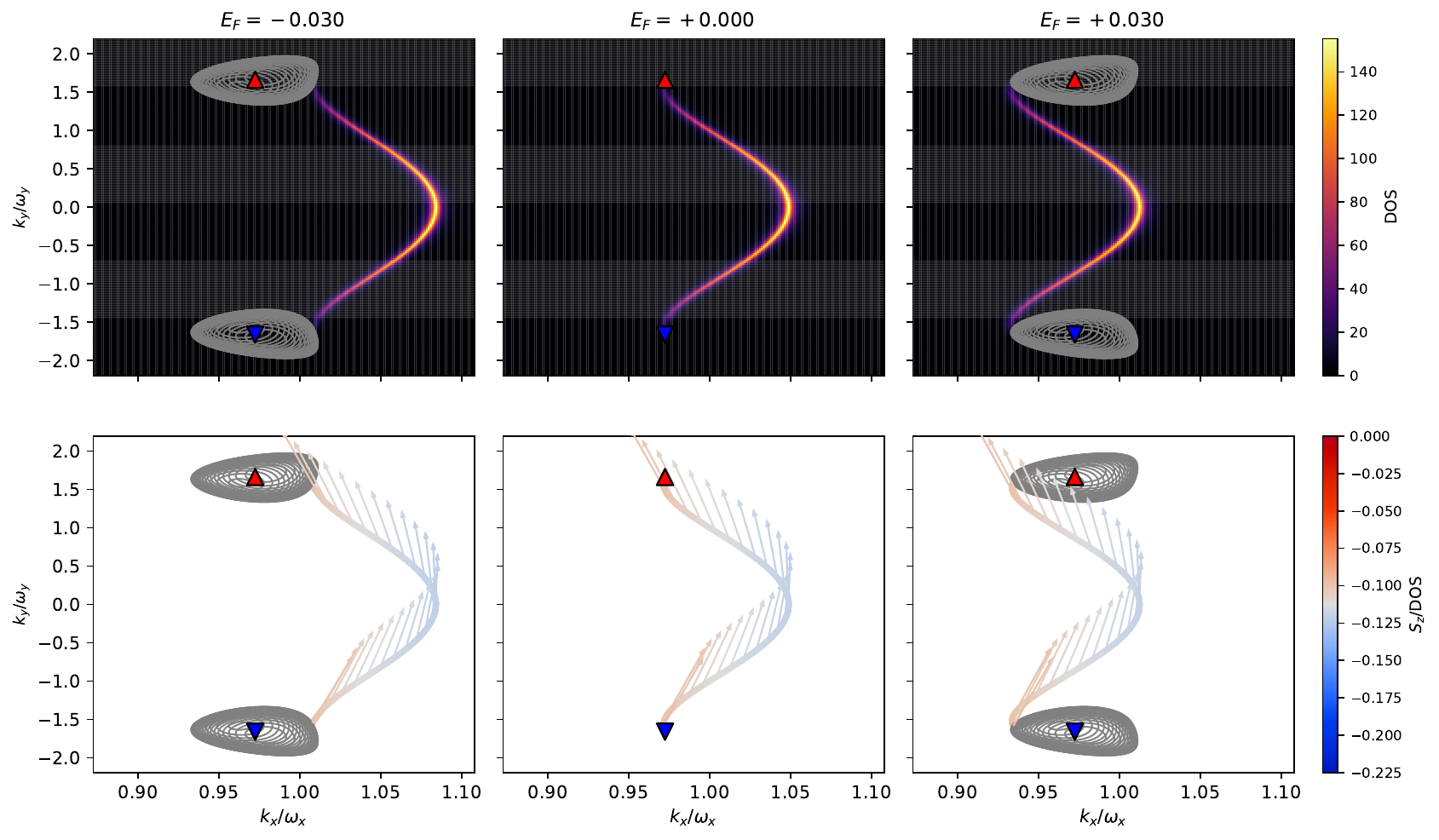}
  \caption{
    The top row shows the surface density of states
    $\rho(k_x, k_y; E_F)$
    at three Fermi energies offset from the Weyl-node energy $E_W$ by
    $\Delta E \in \{-0.03,\, 0,\, +0.03\}$ (left to right).
    The bottom row displays the Fermi arc on the same surface;
    each point is colored by the normalized out-of-plane spin polarization
    $S_z/\rho$, and arrows indicate the in-plane component
    $(S_x, S_y)/\rho$.
    Solid gray curves overlay the bulk constant-energy contours
    $\varepsilon_n(k_x, k_y; k_z) = E_F$ for a representative sampling of
    $k_z \in [-\pi, \pi]$.
    The red $\blacktriangle$ (blue $\blacktriangledown$) marks the surface-Brillouin-zone projections of bulk
    Weyl nodes with positive (negative) chirality.
    The model parameters as in Table~\ref{tab:params}.}
  \label{fig:dos-grid}
\end{figure*}

In order to analyze the Fermi arcs, we consider a finite slab in the z direction. We perform a standard regularization of the continuum model in the $k_z$ direction to construct a tight-binding Hamiltonian (see SM) and calculate the surface density of states and the average value of the spin on the surface. 
Figure \ref{fig:dos-grid} presents the surface density of states in a region spanning a pair of opposite chirality nodes for different values of the Fermi energy: below, at, and above the node energy. 
The solid curves are constant-energy contours of the bulk Hamiltonian,
$\varepsilon_n(k_x, k_y; k_z) = E_F$, at a fixed values of $k_z$;
the family of curves shown samples $k_z \in [-\pi, \pi]$ and together
traces out the projection of the bulk Fermi surface onto the surface
Brillouin zone. The Fermi arc is therefore a genuine
surface mode: it threads the regions free of bulk contours and
terminates at the contours surrounding the Weyl-node projections.
The SOC parameters were chosen here to produce a spin texture on the Fermi arc that resembles the one obtained in {\it ab initio} calculations~\cite{mathisen2026fermiologyspinpolarizationtopological}. Different k-space spin orientations can be obtained depending on the full set of parameters, as shown in the SM.

Considering only the on-site couplings $g_1,g_2$, the projection $s$ along $\hat{\mathbf n}$ is conserved: each eigenstate is a spin eigenstate, with $\langle\vec{\sigma}\rangle=s\,\hat{\mathbf n}$ exactly. The linear-in-$k$ terms break this as their transverse parts do not commute with
$\hat{\mathbf n}\cdot\vec{\sigma}$. As a consequence, $s$ is in general no longer a good quantum number, and the meaningful quantity is the expectation value $\langle\vec{\sigma}\rangle$, the canting being its deviation from $\pm\hat{\mathbf n}$.
The large on-site couplings $g_1,g_2$ lock the unperturbed states to $\pm\hat{\mathbf n}$, so the momentum dependence of the spin direction arises entirely from the transverse ($\perp\hat{\mathbf n}$) parts of the
linear-in-$k$ terms. The longitudinal parts displace the nodes but leave the spin along $\pm\hat{\mathbf n}$. 
The canting direction is set by the spin factor: because $\hat{\mathbf n}$ lies in the $yz$
plane, the two transverse axes are $\hat{\mathbf x}$ (out of plane) and $\hat{\mathbf m}=\hat{\mathbf n}\times\hat{\mathbf x}$ (in plane), so the transverse part of $\sigma_x$ lies along $\hat{\mathbf x}$ while those of $\sigma_y,\sigma_z$ lie along $\hat{\mathbf m}$; the induced canting has its same-axis component along this direction, together with a generically comparable component along the orthogonal transverse axis (see SM). 
The amplitude is set by the orbital factor: since the spin observable $\vec{\sigma}\otimes I_\mathrm{orb}$ is trivial on the orbital sector, $\tau_j$ enters only through its average $\langle\tau_j\rangle$, and the spin cants in proportion to this orbital polarization. 
In each block the orbital pseudospin points along $(f_1,f_2+sg,f_3)$, so its components are $\langle\tau_x\rangle\propto f_1$, $\langle\tau_z\rangle\propto f_3$, and $\langle\tau_0\rangle=1$; the same-axis canting therefore scales as $f_1$ for $\tau_x$ and as $f_3$ for $\tau_z$, while the orthogonal component carries the complementary weights ($f_3$ and $f_1$, respectively; see SM); for $\tau_0$ the canting is
unsuppressed.
On the surface $k_x$ and $k_y$ remain good quantum numbers whereas $k_z$ does not: the arc states are localized in $z$ and are therefore superpositions of $k_z$. The arc spin texture is therefore a weighted $k_z$-average of the bulk texture.

\begin{table}[htbp]
\centering
\caption{Effect of a linear-in-$k$ SOC term $\sigma_i\otimes\tau_j$, separated
into its spin and orbital factors. Relative to the on-site axis
$\hat{\mathbf n}=(0,g_2,g_1)/g$ each term splits into a longitudinal ($\parallel$)
part, which shifts a block field and displaces the node, and a transverse
($\perp$) part, which cants the spin (first order) and renormalises
$g\!\to\!g_{\mathrm{eff}}$ (second order). The spin factor fixes the node-shift
weight $\hat{\mathbf n}\!\cdot\!\hat{\mathbf e}_i$ and the transverse spin axis
($\hat{\mathbf m}=\hat{\mathbf n}\times\hat{\mathbf x}$), along which the
same-axis part of the canting occurs; the orbital factor
fixes the shifted field, the canting amplitude $\langle\tau_j\rangle$, and
whether $g_{\mathrm{eff}}$ is reduced ($<g$) or enhanced ($>g$). The explicit
$g_{\mathrm{eff}}$ is given in the SM.}
\label{tab:soc_effects}
\begin{tabular}{@{}llll@{}}
\toprule
\multicolumn{4}{@{}l}{\textbf{Spin factor}}\\
$\sigma_i$ & $\hat{\mathbf n}\!\cdot\!\hat{\mathbf e}_i$\,($\parallel$) & \multicolumn{2}{l}{transverse spin axis\,($\perp$)}\\
\midrule
$\sigma_x$ & $0$     & \multicolumn{2}{l}{$\hat{\mathbf x}$, out of plane}\\
$\sigma_y$ & $g_2/g$ & \multicolumn{2}{l}{$\hat{\mathbf m}$, in plane}\\
$\sigma_z$ & $g_1/g$ & \multicolumn{2}{l}{$\hat{\mathbf m}$, in plane}\\
\midrule[\heavyrulewidth]
\multicolumn{4}{@{}l}{\textbf{Orbital factor}}\\
$\tau_j$ & field\,($\parallel$) & $\langle\tau_j\rangle$\,($\perp$) & $g_{\mathrm{eff}}$\,($\perp$)\\
\midrule
$\tau_x$ & $f_1$       & $\propto f_1$ & $<g$\\
$\tau_z$ & $f_3$       & $\propto f_3$ & $<g$\\
$\tau_0$ & cone tilt   & unsuppressed  & $>g$\\
\bottomrule
\end{tabular}
\end{table}

\textit{$C_3$ symmetrization.}
The model developed so far describes the low-energy states close to a single
$\Gamma$--$M$ line and is constrained only by the mirror $M_x$ and time
reversal $\mathcal{T}$. The trigonal crystal, however, has point group
$C_{3v}$, whose threefold rotation $C_3$ about the $z$ axis maps this line onto
the two equivalent $\Gamma$--$M$ directions. We restore the full point symmetry
by promoting the model to a multi-sector description built from three rotated
copies.

The rotation acts on momentum as the planar rotation $R\equiv R_{2\pi/3}$ (with
$k_z$ invariant) and on the internal degrees of freedom as
$\mathcal{C}=e^{-i\frac{\pi}{3}\sigma_z}\otimes e^{-i\frac{2\pi}{3}\tau_y}$,
$\mathcal{C}^3=-I_4$, where the orbital factor is the in-plane rotation of
$\{p_x,p_y\}$ (with $\tau_y\sim L_z$) and the spin factor the spin-$\tfrac12$
rotation. The symmetry to be imposed reads
$\mathcal{C}\,H(\mathbf{k})\,\mathcal{C}^{-1}=H(R\mathbf{k})$.

Introducing a sector index $v=0,1,2$ for the three $\Gamma$--$M$ directions and
identifying the model above with the $v=0$ block $H_0$, the remaining blocks
follow by rotation,
\begin{equation}
    H_v(\mathbf{k})=\mathcal{C}^{\,v}\,H_0\!\left(R^{-v}\mathbf{k}\right)
    \mathcal{C}^{-v},
    \label{eq:Hv}
\end{equation}
and assemble into
\begin{equation}
    H_{C_3}(\mathbf{k})=\bigoplus_{v=0}^{2}H_v(\mathbf{k}),
    \label{eq:HC3}
\end{equation}
a $12\times12$ Hamiltonian invariant under
\begin{equation}
    \hat{C}_3=P\otimes\mathcal{C},\qquad
    P=\begin{pmatrix}0&0&1\\1&0&0\\0&1&0\end{pmatrix},
\end{equation}
with $P$ the cyclic shift $v\to v+1$ (mod 3). Using
$\mathcal{C}\,H_v\,\mathcal{C}^{-1}=H_{v+1}(R\mathbf{k})$ (see SM) one obtains
$\hat{C}_3\,H_{C_3}(\mathbf{k})\,\hat{C}_3^{-1}=H_{C_3}(R\mathbf{k})$. The four
Weyl nodes of the single-sector model are thereby replicated onto their two
$C_3$ images, yielding the twelve Weyl points of the trigonal structure.

Within this construction the three sectors are decoupled and the twelve Weyl
nodes are exact. Inter-sector hybridisation is symmetry-allowed and can be
added as a correction generated from a single prototype block $T(\mathbf{k})$
coupling neighbouring sectors, $C_3$-replicated as in Eq.~(\ref{eq:Hv}); $T$
must obey the same time-reversal constraint as the intra-sector blocks together
with a mirror condition relating it to its own $C_3$ image (see SM). Since the sectors are mutually off-resonant at the node momenta,
additional intervalley couplings displace the nodes at the perturbative level
rather than gapping them.

\section{Conclusions}
We have developed a minimal model for PtBi$_2$ whose construction was guided by identifying the minimal orbital–spin structure underlying the bulk electronic states and by introducing an effective sector degree of freedom that groups sets of Weyl nodes related by time-reversal $\mathcal{T}$ and mirror $M_x$ symmetry, providing a natural framework to describe the experimentally relevant Fermi arcs. By enforcing these symmetries, we derived an effective Hamiltonian that qualitatively reproduces the evolution of Weyl nodes observed in \textit{ab initio} calculations as the spin–orbit coupling strength $\lambda$ is varied. Furthermore, we analyzed the spin structure of the surface states, demonstrating that the model can capture the spin-momentum-locking effects on the Fermi arcs. We also extended this construction to include multiple sectors, leading to a $C_{3v}$-symmetric version of the model that reproduces the point symmetry of the lattice. 

This sector structure is a natural starting point to analyze superconducting pairing mechanisms. A zero-momentum, time-reversal-invariant pairing connects
$\mathbf{k}$ to $-\mathbf{k}$, both of which lie on the same $\Gamma$--$M$
line; the pairing is therefore sector diagonal and the Bogoliubov--de Gennes
Hamiltonian splits into three $C_3$-related blocks. A symmetry classification
of the surface order parameter in terms of the irreducible representations of
$C_{3v}$ has been developed in Ref.~\cite{kkqg-ntcz}, and recent ARPES
experiments indicate a nodal superconducting state on the Fermi
arcs~\cite{changdar2025topological}. A detailed analysis of these pairing channels onto
 the present model is left for future work.

\acknowledgments
During the preparation of this work, two large language models  (Anthropic's Claude Fable 5 and Claude Opus 4.8) were used to assist with drafting and editing the manuscript, to generate Python scripts for numerically checking the analytical results, and to format the figures. All generated code and content were reviewed and verified by the authors.
 
\bibliographystyle{eplbib}

\bibliography{PtBi2_cleaned}

\begin{thebibliography}{10}
\expandafter\ifx\csname url\endcsname\relax\def\url#1{\texttt{#1}}\fi

\bibitem{10.1063/5.0272618}
\Name{Qu J., Veyrat A., Büchner B. \and Dufouleur J.} \REVIEW{Applied Physics
  Reviews}{13}{2026}{021332}.

\bibitem{feng_rashba-like_2019}
\Name{Feng Y., Jiang Q., Feng B., Yang M., Xu T., Liu W., Yang X., Arita M.,
  Schwier E.~F., Shimada K., Jeschke H.~O., Thomale R., Shi Y., Wu X., Xiao S.,
  Qiao S. \and He S.} \REVIEW{Nat. Commun.}{10}{2019}{4765}.

\bibitem{Veyrat2023}
\Name{Veyrat A., Labracherie V., Bashlakov D.~L., Caglieris F., Facio J.~I.,
  Shipunov G., Charvin T., Acharya R., Naidyuk Y., Giraud R., van~den Brink J.,
  B\"{u}chner B., Hess C., Aswartham S. \and Dufouleur J.} \REVIEW{Nano
  Lett.}{23}{2023}{1229}.

\bibitem{Kuibarov2023}
\Name{Kuibarov A., Suvorov O., Vocaturo R., Fedorov A., Lou R., Merkwitz L.,
  Voroshnin V., Facio J.~I., Koepernik K., Yaresko A., Shipunov G., Aswartham
  S., van~den Brink J., B\"{u}chner B. \and Borisenko S.}
  \REVIEW{Nature}{626}{2024}{294–299}.

\bibitem{hoffmann2024fermi}
\Name{Hoffmann S., Schimmel S., Vocaturo R., Puig J., Shipunov G., Janson O.,
  Aswartham S., Baumann D., B{\"u}chner B., Brink J. v.~d. \etal} \REVIEW{Adv.
  Phys. Res.}{}{2024}{2400150}.

\bibitem{n5pz-j2sl}
\Name{O'Leary E., Li Z., Wang L.-L., Schrunk B., Eaton A., Canfield P.~C. \and
  Kaminski A.} \REVIEW{Phys. Rev. B}{112}{2025}{085154}.

\bibitem{changdar2025topological}
\Name{Changdar S., Suvorov O., Kuibarov A., Thirupathaiah S., Shipunov G.,
  Aswartham S., Wurmehl S., Kovalchuk I., Koepernik K., Timm C. \etal}
  \REVIEW{Nature}{647}{2025}{613}.

\bibitem{mathisen2026fermiologyspinpolarizationtopological}
\Name{Mathisen A.~C., Tan X.~L., Brinkman S.~S., Mæland K., Göhler F.,
  Øyvind Finnseth, Shipunov G., Pabst F., Lemos M.~A., Thiagarajan B., Polley
  C., Trauzettel B., Isaeva A., Facio J.~I. \and Bentmann H.} \REVIEW{arXiv
  2607.01947}{}{2026}{}.

\bibitem{kuibarov_measuring_2025}
\Name{Kuibarov A., Changdar S., Fedorov A., Lou R., Suvorov O., Misheneva V.,
  Harnagea L., Kovalchuk I., Wurmehl S., Büchner B. \and Borisenko S.}
  \REVIEW{Phys. Rev. B}{112}{2025}{144518} publisher: American Physical
  Society.

\bibitem{PhysRevB.110.054504}
\Name{Vocaturo R., Koepernik K., Facio J.~I., Timm C., Fulga I.~C., Janson O.
  \and van~den Brink J.} \REVIEW{Phys. Rev. B}{110}{2024}{054504}.

\bibitem{PhysRevB.110.125148}
\Name{Zhu A., Chen Z., Han M., Liu X., Chen X., Han Y., Zheng G., Zhu X., Gao
  W. \and Tian M.} \REVIEW{Phys. Rev. B}{110}{2024}{125148}.

\bibitem{PhysRevResearch.7.013025}
\Name{Majchrzak P., Sanders C., Zhang Y., Kuibarov A., Suvorov O., Springate
  E., Kovalchuk I., Aswartham S., Shipunov G., B\"uchner B., Yaresko A.,
  Borisenko S. \and Hofmann P.} \REVIEW{Phys. Rev. Res.}{7}{2025}{013025}.

\bibitem{palumbo2025gapless}
\Name{Palumbo S., Cornaglia P.~S. \and Facio J.~I.} \REVIEW{Physical Review
  B}{112}{2025}{L201117}.

\bibitem{palumbo2025interplay}
\Name{Palumbo S., Cornaglia P.~S. \and Facio J.~I.} \REVIEW{Phys. Rev.
  B}{112}{2025}{205125}.

\bibitem{f66s-m6jy}
\Name{Caglieris F., Ceccardi M., Efremov D., Shipunov G., Kovalchuk I.,
  Aswartham S., Veyrat A., Dufouleur J., Marr\'e D., B\"uchner B. \and Hess C.}
  \REVIEW{Phys. Rev. Mater.}{9}{2025}{084202}.

\bibitem{schimmel2024surface}
\Name{Schimmel S., Fasano Y., Hoffmann S., Besproswanny J., Corredor~Bohorquez
  L.~T., Puig J., Elshalem B.-C., Kalisky B., Shipunov G., Baumann D. \etal}
  \REVIEW{Nat. Commun.}{15}{2024}{9895}.

\bibitem{Zabala_2024}
\Name{Zabala J., Correa V.~F., Castro F.~J. \and Pedrazzini P.} \REVIEW{J.
  Phys. Condens. Matter}{36}{2024}{285701}.

\bibitem{bdtb-mb8c}
\Name{Trama M., K\"onye V., Fulga I.~C. \and van~den Brink J.} \REVIEW{Phys.
  Rev. B}{112}{2025}{064514}.

\bibitem{kkqg-ntcz}
\Name{Waje H., Jakubczyk F., van~den Brink J. \and Timm C.} \REVIEW{Phys. Rev.
  B}{112}{2025}{144519}.

\bibitem{47vs-qgzk}
\Name{M\ae{}land K., Bahari M. \and Trauzettel B.} \REVIEW{Phys. Rev.
  B}{112}{2025}{104507}.

\bibitem{besproswanny2025}
\Name{Besproswanny J., Schimmel S., Fasano Y., Shipunov G., Aswartham S.,
  Baumann D., Büchner B. \and Hess C.} \REVIEW{arXiv 2507.10187}{}{2025}{}.

\bibitem{maeland2025mechanism}
\Name{Mæland K., Sangiovanni G. \and Trauzettel B.} \REVIEW{arXiv
  2512.09994}{}{2025}{}.

\bibitem{buccheri2026}
\Name{Buccheri F., de~Martino A. \and van~den Brink J.} \REVIEW{arXiv
  2606.02371}{}{2026}{}.

\bibitem{vocaturo2026}
\Name{Vocaturo R. \and Trama M.} \REVIEW{arXiv 2604.26859}{}{2026}{}.

\bibitem{ptbi2:kaiser14}
\Name{Kaiser M., Baranov A.~I. \and Ruck M.} \REVIEW{Z. Anorg. Allgem.
  Chem.}{640}{2014}{2742–2746}.

\bibitem{Shipunov2020}
\Name{Shipunov G., Kovalchuk I., Piening B.~R., Labracherie V., Veyrat A., Wolf
  D., Lubk A., Subakti S., Giraud R., Dufouleur J., Shokri S., Caglieris F.,
  Hess C., Efremov D.~V., B\"uchner B. \and Aswartham S.} \REVIEW{Phys. Rev.
  Mater.}{4}{2020}{124202}.

\bibitem{varjas2018qsymm}
\Name{Varjas D., Rosdahl T.~{\"O}. \and Akhmerov A.~R.} \REVIEW{New J.
  Phys.}{20}{2018}{093026}.

\end{thebibliography}

\end{document}